\RequirePackage{fix-cm}
\documentclass[smallcondensed]{svjour3}     % onecolumn (ditto)
\smartqed  % flush right qed marks, e.g. at end of proof
\usepackage{graphicx}
\usepackage{amsmath}
\usepackage{color}
\begin{document}

\title{Towards Sympathetic Cooling of Single (Anti-)Protons}

\author{Teresa Meiners         \and
        Malte Niemann \and 
        Johannes Mielke \and
        Matthias Borchert \and
        Nicolas Pulido \and
        Juan M. Cornejo \and
        Stefan Ulmer \and
        Christian Ospelkaus
}

\institute{T. Meiners \at
              Institut f\"ur Quantenoptik, Leibniz Universit\"at Hannover, Welfengarten 1, 30167 Hannover, Germany \\
              Tel.: +49-511-7624656, 
              Fax: +49-511-7622211, 
              \email{t.meiners@iqo.uni-hannover.de}
           \and
           M. Niemann, J. Mielke, N. Pulido and J. M. Cornejo \at
              Institut f\"ur Quantenoptik, Leibniz Universit\"at Hannover, Welfengarten 1, 30167 Hannover, Germany
           \and
           M. Borchert \at
           Institut f\"ur Quantenoptik, Leibniz Universit\"at Hannover, Welfengarten 1, 30167 Hannover, Germany \at
           RIKEN, Ulmer Fundamental Symmetries Laboratory, Wako, Saitama 351-0198, Japan
           \and
			S. Ulmer \at
           RIKEN, Ulmer Fundamental Symmetries Laboratory, Wako, Saitama 351-0198, Japan
           \and
           C. Ospelkaus \at
           Institut f\"ur Quantenoptik, Leibniz Universit\"at Hannover, Welfengarten 1, 30167 Hannover, Germany \at
           Physikalisch Technische Bundesanstalt, Bundesallee 100, 38116 Braunschweig, Germany
}

\date{Received: date / Accepted: date}

\maketitle

\begin{abstract}
We present methods to manipulate and detect the motional state and the spin state of a single antiproton or proton which are currently under development within the BASE (Baryon Antibaryon Symmetry Experiment) collaboration. These methods include sympathetic laser cooling of a single (anti-)proton using a co-trapped atomic ion as well as quantum logic spectroscopy with the two particles and could be implemented within the collaboration for state preparation and state readout in the antiproton $g$-factor measurement experiment at CERN. In our project, these techniques shall be applied using a single $^9\text{Be}^+$ ion as the atomic ion in a Penning trap system at a magnetic field of 5~T. As an intermediate step, a controlled interaction of two beryllium ions in a double-well potential as well as sympathetic cooling of one ion by the other shall be demonstrated.
\keywords{Penning traps \and laser cooling \and motional coupling \and atomic ion \and $g$-factor \and antiproton}
\end{abstract}

\section{Introduction}
\label{intro}
Comparing the magnetic moment of the proton and the antiproton is a stringent test of CPT invariance \cite{ding2016}. The $g$-factor of the proton has been determined with a precision of 0.3 ppb \cite{schneider2017} and that of the anti-proton with a precision of 1.5 ppb \cite{smorra2017}. The $g$-factor $g$ can be determined by measuring the free cyclotron frequency $\omega_\text{C}$ and the Larmor frequency $\omega_\text{L}$, because $g=2\frac{\omega_\text{L}}{\omega_\text{C}}$. These frequency measurements are performed in Penning traps. The free cyclotron frequency is commonly determined using image current detection and the Larmor frequency by applying the continuous Stern Gerlach effect \cite{dehmelt1973}. An alternative method of determining the Larmor frequency following a proposal by Heinzen and Wineland \cite{heinzen1990} that enables full control of all motional degrees of freedom of the (anti-)proton is described subsequently. Related ideas could also be implemented using antihydrogen ions cooled by $^9\text{Be}^+$ \cite{hilico2014}.

\section{State preparation and readout of a single (anti-)proton}
The experimental sequence described in the following will be carried out in a trap stack composed of four different Penning traps: A laser cooling and detection trap, a Coulomb coupling trap, a spin-motion coupling trap, and a precision trap \cite{niemann2014}, \cite{meiners2017}.

For state preparation the $^9\text{Be}^+$ ion and the (anti-)proton are initialized to their motional ground states and at the same time the $^9\text{Be}^+$ ion is prepared in the ``spin down'' state of two defined (hyperfine) qubit states. For the $^9\text{Be}^+$ ion both is achieved by using laser Doppler cooling, followed by laser sideband cooling \cite{segal2012}. The (anti-)proton ground state is prepared by applying sympathetic cooling via the remote Coulomb interaction in a double-well potential through the beryllium ion. Afterwards, the initial internal spin state of the \mbox{(anti-)proton} is mapped to the motional state with a radio-frequency sideband pulse while at the same time preparing the spin in a known state, depending on the type (red or blue) of sideband used. This motional state is then transferred to the $^9\text{Be}^+$ ion via Coulomb interaction in the double-well potential \cite{smorra2015} and mapped to one of the qubit states by applying a motional sideband pulse. Read-out of the qubit state using fluorescence detection reveals the initial spin state of the (anti-)proton \cite{meiners2016}.

To determine the Larmor frequency of the (anti-)proton the particle is irradiated with radio frequency pulses at different frequencies close to the Larmor frequency to drive a spin flip. After each attempt the spin state of the \mbox{(anti-)proton} is determined as described above. This measurement sequence results in a probability distribution from which the Larmor frequency can be calculated.

\section{Current experimental setup for motional coupling of two ions}
With our current experimental setup we are aiming for demonstration of motional coupling of two beryllium ions in a Penning trap.
The main part of the experimental setup is a cylindrical Penning trap stack that consists of three individual Penning traps. This stack is fixed to a support structure placed inside a vacuum chamber and mounted to a superconducting magnet and a mechanical cryo-cooler. The magnet provides a homogeneous magnetic field of 5 Tesla. The mechanical cryo-cooler provides two stages, one at a temperature near that of liquid helium to cool the trap stack and another at a temperature near that of liquid nitrogen for shielding the trap stack from thermal radiation.

The trap stack currently placed on the system contains the following individual Penning traps: a laser trap, a coupling trap, and a storage trap. With this system Doppler cooling of a single $^9\text{Be}^+$ ion can be achieved as well as sympathetic cooling of a second $^9\text{Be}^+$ ion via Coulomb interaction in the double-well potential. 

For the purpose of Doppler cooling, the laser trap has been designed and built. This trap has optical access for production, manipulation, and imaging of the ion. For production, a pulsed laser  operating at 1064 nm and a continuous wave (cw) laser system operating at 235 nm shall be used for ablation and photoionization of beryllium atoms, respectively. For Doppler cooling, repumping, and fluorescence detection a cw laser system operating at 313 nm will be used. Its output beam crosses the trap center at an angle of $45^\circ$ with respect to the magnetic field direction. The description of the cw laser systems can be found in \cite{meiners2017} and \cite{hannig2018}.

For coupling of the axial modes of motion of two beryllium ions a double-well potential has been engineered in the coupling trap. The axial trap frequency of the two ions, which needs to be equal to achieve efficient energy exchange, has been calculated to be 129~kHz for an ion-to-ion distance of 1.24~mm. Approximating the double-well potential with the potentials of two quantized harmonic oscillators, the energy exchange rate for the motional energy of the ions can be derived \cite{brown2011}:

\begin{equation}
\Omega_\text{ex} = \frac{q^2}{4 \pi \epsilon_0 s_0^3 m \omega_0}
\end{equation}

where $q$ is the ion's charge, $\epsilon_0$ the permittivity of free space, $s_0$ the ion-to-ion distance, $m$ the ion's mass, and $\omega_0$ the ion's axial trap frequency. In our case the exchange rate is 10~s$^{-1}$ which results in an exchange time $t_\text{ex}=\frac{\pi}{2 \Omega_\text{ex}}=$157~ms to swap the motional states of the two ions. 

\section{Summary and prospects}
We have described the goals and the current status of an experimental setup to demonstrate quantum logic spectroscopy of (anti-)protons. As a first step, sympathetic cooling of one $^9\text{Be}^+$ ion by another one can be demonstrated using their Coulomb interaction in a double-well potential. In the future this techique shall be applied to single (anti-)protons with a single beryllium ion in its motional ground state prior to carrying out quantum logic spectroscopy. 

In order to start experiments with protons, we have already designed and produced a proton source based on electron bombardment of an organic target material.

\begin{acknowledgements}
We acknowledge funding from QUEST, LUH, PTB, ERC StG ``QLEDS", and DFG through SFB 1227 ``DQ-\emph{mat}".
	
We acknowledge support by Tobias Florin and Julian Pick in building laser systems and designing a proton source.

\end{acknowledgements}


\begin{thebibliography}{}

\bibitem{ding2016}
Y. Ding and V. A. Kosteleck\'y, Lorentz-violating spinor electrodynamics and Penning traps, Phys. Rev. D, 94, 056008 (2016)

\bibitem{schneider2017}
G. Schneider \textit{et al.}, Double-trap measurement of the proton magnetic moment at 0.3 parts per billion precision, Science, 358, 1081-1084 (2017)

\bibitem{smorra2017}
C. Smorra \textit{et al.}, A parts-per-billion measurement of the antiproton magnetic moment, Nature, 550, 371-374 (2017)

\bibitem{dehmelt1973}
H. Dehmelt and P. Ekstr\"om, Proposed g-2/$\delta \omega_z$ Experiment on Single Stored Electron or Positron, Bull. Am. Phys. Soc, 18, 727 (1973)

\bibitem{heinzen1990}
D. J. Heinzen and D. J. Wineland, Quantum-limited cooling and detection of radio-frequency oscillations by laser-cooled ions, Phys. Rev. A, 42, 2977-2994 (1990)

\bibitem{hilico2014}
L. Hilico \textit{et al.}, Preparing single ultra-cold antihydrogen atoms for free-fall in GBAR, International Journal of Modern Physics: Conference Series, 30, 1460269 (2014)

\bibitem{niemann2014}
M. Niemann \textit{et al.}, Proceedings of the Sixth Meeting on CPT and Lorentz Symmetry, 41-44. World Scientific, Singapore (2014)

\bibitem{meiners2017}
T. Meiners \textit{et al.}, Towards Quantum Logic Inspired Cooling and Detection
for Single \mbox{(Anti-)Protons}, JPS Conf. Proc., 18, 011006 (2017)

\bibitem{segal2012}
D. M. Segal and C. Wunderlich, Physics with trapped charged particles, 48-69. Imperial College Press, London (2014)

\bibitem{smorra2015}
Smorra et al., BASE – The Baryon Antibaryon Symmetry Experiment, Eur. Phys. J. Special Topics 224, 3055-3108 (2015)

\bibitem{meiners2016}
T. Meiners \textit{et al.}, Proceedings of the Seventh Meeting on CPT and Lorentz Symmetry, 85-88. World Scientific, Singapore (2017)

\bibitem{hannig2018}
S. Hannig \textit{et al.}, A highly stable monolithic enhancement cavity for second harmonic generation in the ultraviolet, Review of Scientific Instruments, 89, 013106 (2018)

\bibitem{brown2011}
K. R. Brown \textit{et al.}, Coupled quantized mechanical oscillators, Nature, 471, 196-199 (2011)

\end{thebibliography}
\end{document}